\documentclass[aps,a4paper]{revtex4}

\usepackage{graphicx}
\usepackage{amsmath,amssymb,color}
\usepackage[english]{babel}

\parskip=\medskipamount

\newcommand{\eq}[1]{(\ref{#1})}
\newcommand{\fig}[1]{Fig.\ref{#1}}

\newcommand{\be}{\begin{equation}}
\newcommand{\ee}{\end{equation}}

\newcommand{\barr}{\begin{array}}
\newcommand{\earr}{\end{array}}

\newcommand{\beqn}{\begin{eqnarray}}
\newcommand{\eeqn}{\end{eqnarray}}

\newcommand{\bs}{\begin{subequations}}
\newcommand{\es}{\end{subequations}}

\newcommand{\bw}{\begin{widetext}}
\newcommand{\ew}{\end{widetext}}

\newcommand\disp{\displaystyle}

\newcommand{\ii}{\mathrm{i}}

\newcommand{\la}{\left<}
\newcommand{\ra}{\right>}

\newcommand{\ve}{\mathbf}

\begin{document}

\title{Many-body contacts in fractal polymer chains and fBm trajectories}

\author{K.E. Polovnikov$^{1,2}$, S. Nechaev$^{3,4}$ and M.V. Tamm$^{2,5}$}

\affiliation{$^1$ Skolkovo Institute of Science and Technology, 143026 Skolkovo, Russia \\ $^2$ Faculty of Physics, Lomonosov Moscow State University, 119992 Moscow, Russia \\ $^3$Interdisciplinary Scientific Center Poncelet (ISCP), 119002, Moscow, Russia \\ $^4$ Lebedev Physical Institute RAS, 119991, Moscow, Russia \\ $^5$ Department of Applied Mathematics, MIEM, National Research University Higher School of Economics, 101000, Moscow, Russia}

\date{\today}

\begin{abstract}

We calculate the probabilities that a trajectory of a fractional Brownian motion with arbitrary fractal dimension $d_f$ visits the same spot $n \geq 3$ times, at given moments $t_1, ..., t_n$, and obtain a determinant expression for these probabilities in terms of a displacement-displacement covariance matrix. Except for the standard Brownian trajectories with $d_f = 2$, the resulting many-body contact probabilities cannot be factorized into a product of single loop contributions. Within a Gaussian network model of a self-interacting polymer chain, which we suggested recently, the probabilities we calculate here can be interpreted as probabilities of multi-body contacts in a fractal polymer conformation with the same fractal dimension $d_f$. This Gaussian approach, which implies a mapping from fractional Brownian motion trajectories to polymer conformations, can be used as a semiquantitative model of polymer chains in topologically-stabilized conformations, e.g., in melts of unconcatenated rings or in the chromatin fiber, which is the material medium containing genetic information. The model presented here can be used, therefore, as a benchmark for interpretation of the data of many-body contacts in genomes, which we expect to be available soon in, e.g., Hi-C experiments.

\end{abstract}
\maketitle


What is the probability for a Brownian trajectory to visit the same space point several times? What is the number of points in space visited exactly $n$ times by a single polymer chain? These questions are of significant importance for the classical polymer physics \cite{deGennes_book,DoiEdwards, GrosbergKhokhlov, Rubinstein}. Indeed, it is known that conformations of ideal polymer chains and Brownian trajectories are isomorphic up to the redefinition of time: the time for a Brownian motion is replaced by the contour length for a polymer. As a result, the points in space visited $n$ times by a Brownian trajectory correspond exactly to the many-body interactions in a polymer chain. Answers to the above questions (provided that one allows for a proper coarse-graining to avoid ultraviolet divergencies), are essentially quite simple: since a Brownian motion is a Markovian process, every loop it forms is independent from the other. As a result, the probability of a "rosette", $P_n(t_1,...,t_n)$, for the trajectory to return to a given point at sequential time moments $t_1<t_2< ... <t_n$ is equal to $\prod_{i=1}^{n-1} P_2(t_{j+1}-t_j)$, where $P_2(t_{j+1}-t_j)$ is the two-body contact probability. Interestingly, even in this case the total number of points visited exactly $n$ times exhibits rich and interesting scaling behavior depending on $n$ and $d$\cite{khokhlov_notes, maj_tamm, turban}. However, this simple factorization does not hold anymore for the case of fractional Brownian motion with $d_f\neq 2$. The increments of a fractional Brownian motion have long-ranged slowly decaying correlations, and the probability of, say, second loop will essentially depend on the fact that the first loop is formed and on its length. In this note we study these loop correlations and explicitly calculate the resulting multiloop probabilities for fractal polymers with $d_f>2$.

Apart from the fundamental interest of our results for the study of general properties of fractional Brownian motion and for polymer physics, the probabilities of multi-body interactions are of significant importance for biophysical applications. Recently in \cite{polovnikov_sm18} we proposed to use the fractional Brownian motion trajectories as an analytically simple semiquantitative model of collapsed polymer conformations with fractal dimension $d_f>2$, including polymer conformations in topologically stabilized states, such as melts of nonconcatenated rings and chromosomes in living nuclei. Technically, our computation of the multiloop probabilities is based on a specific way of constructing ensembles of discretized fractional Brownian trajectories based on a Gibbs sampling with the quadratic Hamiltonian suggested in \cite{polovnikov_sm18}. In the framework of our model it is possible to calculate different experimentally accessible static \cite{polovnikov_sm18} and dynamic \cite{holcman, polovnikov_prl18} properties of collapsed fractal chains with a given fractal dimension.


Modern experimental techniques of genome-wide chromosome conformation capture (Hi-C)\cite{dekker,lieberman-eiden} allow to describe a conformation of the chromatin (the constituting fiber of the chromosome, containing dsDNA with genetic information, as well as various proteins attached to it) by tabulating a full collection of \emph{in vivo} pairwise contacts of the chromatin fiber with itself. This Hi-C data has proven hugely beneficial for our understanding of chromosome conformations under different conditions. There exist a wide range of theories used to rationalize the existing Hi-C data \cite{gns,grosberg93,mirny11,grosberg_review,nicodemi,rosa14,shin14,loop_extrusion,loop_extrusion2}, however all these models seem to agree that in a wide range of length scales the resulting equilibrium chromatin packing is approximately fractal with transient fractal dimension $d_f$ lying in the interval $2\leq d_f \leq 4$, $d_f =3$, which corresponds to a space-filling curve, being the most obvious candidate for the true limiting fractal dimension in the limit of infinitely long chains.

Additional evidence concerning chromatin conformations should come from consideration of \emph{triple} and many-body contacts, which, generally speaking, contain information on the properties of the fiber packing, which is not reducible to the information obtained from two-loci contacts. Recent advances in experimental techniques \cite{pedro16,oudelaar18} allow to expect that soon it will be possible to collect enough statistics on triple contacts from the Hi-C experiments. Therefore, theoretical approaches allowing to make sense of this upcoming data are urgently needed. Here we calculate the three-body and many-body contact probabilities in the simple model of a Gaussian polymer chain, which can be used as a basic benchmark to compare the experiment data with. In particular, we propose here to measure experimentally the two-loop correlation factor (the ratio of a three-body contact probability to the product of the probabilities of two independent loops of the same size) as an important characteristic of the chromatin conformation statistics, and provide concrete predictions for the value of this factor within our fractal Gaussian polymer model.

To begin with, define the trajectory of a discretized random walk, or, similarly, a conformation of a polymer chain by a set of coordinates $\{\ve X\} = \{\ve x_0, \ve x_1, ..., \ve x_N\}$. In the random walk terms, $\{\ve X\}$ characterizes the position of the walker at sequential time ticks, $t = 0,1,.., N$. In polymer terms, fixing the Gaussian transitions between sequential "times" in the set $\{\ve X\}$, we define the standard beads-on-string model \cite{GrosbergKhokhlov}. In what follows we assume that $N \gg 1$, and the process is stationary in a sense that mean distance between $\ve x_j$ and $\ve x_{j+1}$ is $j$-independent:
\be
\la (\ve x_{j+1} - \ve x_j)^2 \ra = b^2.
\ee
There exist a unique stationary measure $P(\ve X)$ over the realizations of the walk $\ve X$, which satisfies
\be
p(({\ve x}_k - \ve x_m) = \ve y) = \int d\ve X P(\ve X) \delta({\ve x}_k - \ve x_m - \ve y)= \left(\frac{3}{2\pi b^2 s^{2/d_f}}\right)^{3/2} \exp\left(-\frac{3 \ve y^2}{2 b^2 s^{2/d_f}}\right); \quad s = |k-m|
\label{fGaussian}
\ee
for all $k, m$ and some fixed constant $d_f$ called fractal dimension of the walk. The variable $s$ in \eq{fGaussian} plays a role of the time between moments $k$ and $m$ in the random walk language, while for a polymer it is the contour distance along the chain. The process described by the stationary measure $P(\ve X)$ is called fractional Brownian motion. There exist numerous equivalent definitions of this process \cite{mandelbrot}, and it is often characterized by the so-called Hurst exponent $H = 1/d_f$. Clearly, the value $d_f = 2$ ($H = 1/2$), which corresponds to Brownian motion, is the only case when the process described by \eq{fGaussian} is Markovian. Indeed, \eq{fGaussian} does not respect the Chapman-Kolmogorov relation for any $d_f \neq 2$. It follows immediately from \eq{fGaussian} that the mean-square spatial distance between two points on a trajectory grows algebraically with $s$
\be
\langle\left(\mathbf{x}_k-\mathbf{x}_m\right)^2\rangle =b^2 s^{2/d_f}
\label{fd}
\ee
and the covariance
\be
\la (\ve x_k - \ve x_m)(\ve x_n - \ve x_m) \ra = \frac{b^2}{2}\left(s^{2/d_f} + (s')^{2/d_f} - |s-s'|^{2/d_f}\right)
\label{fd2}
\ee
where we used notation $s' =|n-m|$.

It has been shown recently in \cite{polovnikov_sm18} that in the limit of large $N$ one can approximate $P(\ve X)$ for the subdiffusive ($d_f>2$) fractal Brownian motion (fBm) by a Gibbs measure with a simple quadratic Hamiltonian:
\be
P(\ve X) = \exp (-V (\ve X)); \quad V(\ve X) = \sum_{i<j} A_{ij} \left(\mathbf{x}_{i}-\mathbf{x}_{j}\right)^2
\label{quadr}
\ee
with a proper choice of interactions coefficients $A_{ij}$ (here and below we use lowercase and uppercase bold letters to denotes vectors in 3-dimensional and 3N-dimensional space, respectively). In particular, if the coefficients $A_{km}$ depend only on the chemical distance between monomers $|k-m|=s$, so that $A_{km}=A(s)$, and if for $s\gg 1$ $A(s)$ decays algebraically:
\be
A(s) \sim \disp c\, s^{-\gamma}
\label{decay}
\ee
with some $c>0$, then depending on $\gamma$ there are three possible asymptotic regimes of polymer chain statistics.
\begin{enumerate}

\item If $\gamma \leq 2$ all monomers (points of the trajectory) asymptotically merge, and \be \langle\left(\mathbf{x}_k-\mathbf{x}_m\right)^2\rangle \to 0 \quad \mbox{when $N \to \infty$}
\label{fd1}
\ee
regardless $s=|k-m|$;
\item If $\gamma>3$ the interaction is irrelevant and the large scale properties of the trajectory are indistinguishable from the standard Brownian motion with $d_f = 2$.
\item Finally, and most interestingly, if $2<\gamma <3$ the relation \eq{fGaussian} holds for $1 \ll s \ll N$ with some renormalized $\bar b (c,\gamma)$ and a non-trivial fractal dimension \be
d_f = \frac{2}{\gamma-2}
\label{df}
\ee
\end{enumerate}
The value of $\gamma = 3$ is critical, giving rise to the logarithmic corrections to \eq{fGaussian}.

\begin{figure}[ht]
\centerline{\includegraphics[width=16cm]{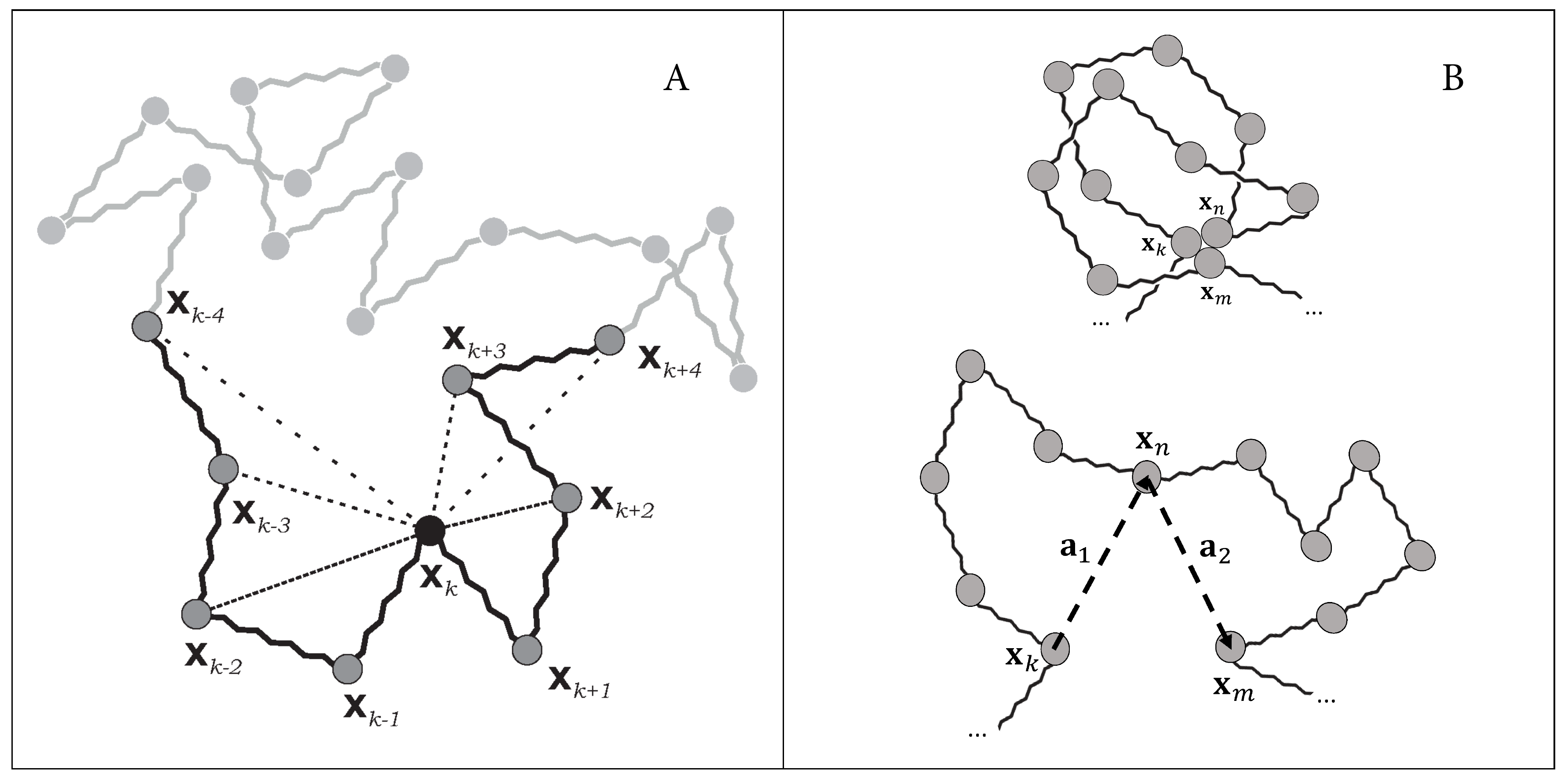}}
\caption{A: Schematic image of the pairwise interactions \eq{quadr} $V_{\ve x_k, \ve x_{m}}$ of the $k$-th monomer ($\mathbf{x}_k$) with adjacent monomers of the chain with coordinates $\mathbf{x}_{k\pm 1},\mathbf{x}_{k\pm 2},\mathbf{x}_{k\pm 3},...$. Elastic constants $a_{km}$ decay algebraically which is depicted by dashed lines with increasing spacing. B (Up): A triple contact, made by monomers $\mathbf{x}_k$, $\mathbf{x}_n$, $\mathbf{x}_m$. Paths $k-n$ and $n-m$ are dependent and the effective volume of the two loops
is less than the product of volumes, occupied by the loops separately. B (Bottom): Typical conformation of the segment.}
\label{fig:01}
\end{figure}

In the polymer context, quadratic interactions in \eq{quadr} can be interpreted as a set of harmonic springs of varying rigidity connecting each pair of monomers, as shown in \fig{fig:01}A. The Hamiltonian of type \eq{quadr} has appeared in the study of 3D structures of proteins \cite{bahar97,haliloglu97,min}, it was used for the description of static and dynamic properties of marginally compact trees with various fractal architectures \cite{dolgushev1,dolgushev2}, as well as for phenomenological study of the Rouse dynamics of non-ideal polymer chains \cite{holcman, polovnikov_prl18}. A proxy Hamiltonian of the same type has been recently used to obtain three-dimensional structures of chromosomes based on experimental Hi-C contact maps \cite{letreut18}. A related hierarchical variational approach for an account of volume interactions of swollen polymer chains had been proposed in \cite{burlatsky}.

Now, we are interested in the probability of forming an $n$-body contact, or, in random walk terms, in the probability for fBm to visit the same spatial point $n$ times. In the simplest case of $n=2$, the probability for two beads separated by chemical distance $s$ to be at the same space point, is
\be
P_2 (s) = p(({\ve x}_k - \ve x_m)=\ve 0) = \left(\frac{3}{2\pi b^2 s^{2/d_f}}\right)^{3/2} = \left(\frac{3}{2\pi \langle\left(\mathbf{x}_k-\mathbf{x}_m\right)^2\rangle}\right)^{3/2}
\label{loop}
\ee
This result, up to the numerical constant, can be easily estimated from the following qualitative argument. If some monomer $k$ is fixed at a given point in space, the monomer $m$ is able to explore the volume of order
\be
V (s) \sim \langle\left(\mathbf{x}_k-\mathbf{x}_m\right)^2\rangle^{3/2}.
\label{twoc}
\ee
If the probabilities to visit all points of this volume are the same, we expect (up to a numerical constant) for the looping probability the expression $P_2(s) \sim V(s)^{-1}$ in a full agreement with \eq{loop}.

Standard Brownian motion is a Markovian process. Accordingly, the probability of forming many body contacts can be constructed as a product of probabilities to form consecutive loops. In particular, the probability to form a "two-loop rosette" with loops of contour lengths $s_1,s_2$ is
\be
P_3^{\text{BM}} (s_1,s_2) = P_2^{\text{BM}}(s_1) P_2 ^{\text{BM}}(s_2)
\label{2-rosette}
\ee
The relation \eq{2-rosette} no longer holds in the non-Markovian case of fBm. Thus, one expects
\be
P_3 (s_1,s_2) = P_2(s_1) P_2 (s_2) f_2(s_1,s_2),
\label{triple}
\ee
where $P_2(s_{1,2})$ are given by \eq{loop}, and $f_2(s_1,s_2)$ is some yet unknown correlation function. In what follows we calculate function $f_2(s_1,s_2)$ explicitly.


Triple contacts are formed by two consecutive loops, which are, typically, correlated due to the long-range memory characteristic of fractional Brownian motion. Hence, the decomposition of the probability into the product of two probabilities of pairwise contacts \eq{2-rosette} is not valid anymore. One can, however, write an explicit expression for $P_3$ by integrating over Gibbs measure with the effective Hamiltonian \eq{quadr}
\be
\displaystyle P_3(k,n,m) = \dfrac{\displaystyle \int P(\ve X) \delta(\ve x_k - \ve x_n) \delta(\ve x_n - \ve x_m)D\ve X}{\displaystyle \int P(\ve X) D\ve X} = \frac{1}{(2\pi)^6} \int d\mathbf{q} \int d\mathbf{q'}\; G(\mathbf{q}, \mathbf{q'}, k, n, m)
\label{p2}
\ee
where we assume without loss of generality that $k<n<m$, the Gibbs distribution $P(\ve X)$ is given by \eq{quadr}, and the Green function, $G(\mathbf{q}, \mathbf{q'}, k, n, m)$, reads
\be
G(\mathbf{q}, \mathbf{q'}, k, n, m) = \frac{1}{Z_N}\int \prod_{i=1}^{N} d \mathbf{x}_i \exp\left(-\frac{1}{2} \sum_{i, j=1}^{N} a_{ij} \mathbf{x}_i \mathbf{x}_j + \ii \mathbf{q} \left(\mathbf{x}_k-\mathbf{x}_n\right) + \ii \mathbf{q'} \left(\mathbf{x}_n-\mathbf{x}_m\right)\right)
\label{two_cont}
\ee
where $Z_N$ is the full partition function of the chain with one link fixed at a given spatial position
\be
Z_N = \int P(\ve X) \delta(\ve x_0) D\ve X = \int \prod_{i=1}^{N} d\mathbf{x}_i \exp\left(-\frac{1}{2}\sum_{i, j=0}^{N} a_{ij} \mathbf{x}_i \mathbf{x}_j \right),
\ee
the matrix $\mathbb{A} = ||a_{ij}||$ is related to the coefficients of the Hamiltonian \eq{quadr} by
\be
a_{ij} = \begin{cases}
\medskip - 4A_{ij}, \quad i\neq j \\
2\displaystyle \sum_j A_{ij} \quad i = j
\end{cases}
\ee
and we used the Fourier transform of the $\delta$-function to obtain the last equality of \eq{p2}. The Green function \eq{two_cont} is an $N$-dimensional Gaussian integral with the linear term $\ve B \ve X$, where components of the vector $\ve B$ are
\be
\mathbf{b}_j = \ii \ve q (\delta_{jk} - \delta_{jn}) + \ii \ve q' (\delta_{jn} - \delta_{jm}).
\label{b}
\ee
It equals therefore
\be
G(\mathbf{q}, \mathbf{q'}, k, n, m) = \exp \left(\frac{1}{2}\mathbf{B}^{T} \mathbb{A}^{-1} \mathbf{B}\right).
\label{g}
\ee
Introduce $\ve A_p$ and $\omega_p$ ($p = 1,2,...N$) -- the eigenvectors and, respectively, eigenvalues of the interaction matrix $\mathbb{A}$, ordered from the smallest eigenvalue to the largest, and scalar (inner) products
\be
\pmb{\beta}_p = \left\langle\ve B | \ve A_p\right\rangle;
\label{eig}
\ee
The quadratic form in \eq{g} can be then rewritten as
\be
\log G(\mathbf{q}, \mathbf{q'}, k, n, m) = \frac{1}{2}\sum_{m=1}^{N} \langle \ve A_m |\pmb \beta_m^{*} \sum_{p=1}^{N} \omega_p^{-1} \pmb \beta_p |\ve A_p \rangle = \frac{1}{2} \sum_{m=1}^{N} \sum_{p=1}^{N} \pmb \beta_m^{*} \omega_p^{-1} \pmb \beta_p \langle \ve A_m |\ve A_p \rangle = \frac{1}{2}\sum_{p=1}^{N} \omega_p^{-1} \left|\pmb \beta_p\right|^2
\label{diag_potential}
\ee
Substituting \eq{b} into \eq{eig} one can express $\beta$'s in terms of the components of the eigenvectors $\ve A$, giving
\be
P_3(k,n,m)=\frac{1}{(2\pi)^{6}} \int d\mathbf{q} \int d\mathbf{q'} \exp\left(-\frac{1}{2} \sum_{p=1}^{N} \omega_p^{-1} \left|\mathbf{q} \left(a_p^k- a_p^n\right)+\mathbf{q}'\left(a_p^n- a_p^m\right)\right|^2\right),
\label{two_cont2}
\ee
where $a_p^i$ denotes the $i$-th component of the vector $\ve A_p$. The right hand side \eq{two_cont2} is a double Gaussian integral, so it can be calculated explicitly:
\be
P_3(k,n,m) = \frac{1}{(2\pi)^3} \frac{1}{(\det \Sigma)^{3/2}}
\label{sigma}
\ee
where elements of the matrix $\Sigma$ are
\be
\sigma_{11} = \sum_{p=1}^{N} \omega_p^{-1} \left|a_p^k - a_p^n\right|^2; \; \; \sigma_{22} = \sum_{p=1}^{N} \omega_p^{-1} \left|a_p^n - a_p^m\right|^2; \;\;
\sigma_{12} = \sigma_{21} = \sum_{p=1}^{N} \omega_p^{-1} \left(a_p^k - a_p^n\right)\left(a_p^n - a_p^m\right);
\ee
and the power 3/2 in the right hand side of \eq{sigma} comes from the fact that $\ve q, \ve q'$ are 3-dimensional vectors. Interestingly, the elements of matrix $\Sigma$ have the meaning of the mean-square distances between corresponding monomers. Indeed, expressing these distances as sums over normal coordinates $\ve u_p = \left\langle\ve X | \ve A_p\right\rangle$ and using the equipartition theorem, which determines the average amplitudes of the normal modes at equilibrium (see \cite{polovnikov_sm18} for more details), one gets $\langle \ve u_p \ve u_s \rangle = 3 \omega_p^{-1} \delta_{ps}$ and, therefore,
\be
\langle \left(\ve x_k - \ve x_n\right)^2 \rangle = 3\sigma_{11}; \; \langle \left(\ve x_n - \ve x_m\right)^2 \rangle = 3\sigma_{22}; \langle \left(\ve x_k - \ve x_n\right)\left(\ve x_n - \ve x_m\right) \rangle = 3\sigma_{12}.
\ee
This correspondence allows to write down the probability of a triple contact in the form suggested in \eq{triple}:
\be
P_3 (k,n,m) = P_2(k,n) P_2 (n,m) f_2(k,n,m),
\label{triple_1}
\ee
where $P_2(k,n)$ is the probability of a double contact \eq{loop}, and the correlator $f_2(k,n,m)$ reads
\be
f_2(k,n,m) = (1 - r^2)^{-3/2}; \;\; r(k,n,m) = \frac{\langle \left(\ve x_k - \ve x_n\right)\left(\ve x_n - \ve x_m\right) \rangle}{\sqrt{\langle \left(\ve x_k - \ve x_n\right)^2 \rangle \langle \left(\ve x_n - \ve x_m\right)^2 \rangle }}
\label{triple_f}
\ee
This result is valid for any Gaussian polymer chain with interactions \eq{twoc}. In particular, for ideal chain (ordinary Brownian motion) vectors $\left(\ve x_k - \ve x_n\right)$ and $\left(\ve x_n - \ve x_m\right)$ are uncorrelated, thus $r = 0$, and one returns to the expression \eq{2-rosette}.

\begin{figure}[ht]
\centerline{\includegraphics[width=16cm]{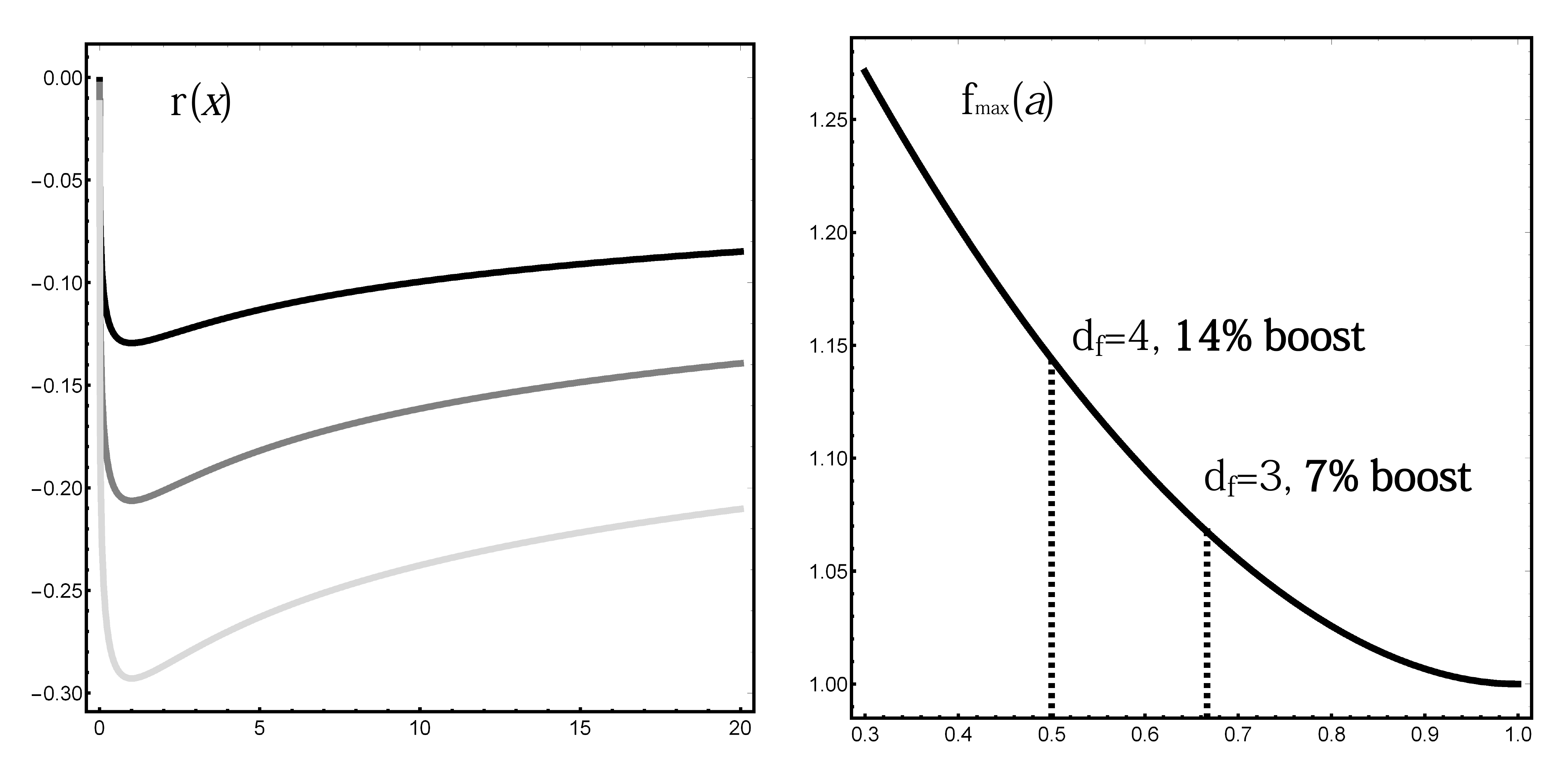}}
\caption{Left: Correlation coefficient $r(\chi)$ between two loops of lengths $s_1$ and $s_2$ as a function of the ratio of their sizes $\chi=s_1/s_2$ as given by \eq{rt} for various values of $d_f$: 2.5 (black), 3 (gray), 4 (light gray); Right: The maximal boost of the two-loops probability $f_{max}(\alpha)$ for the case of equal loops sizes $s_1 = s_2$ as compared to loops being independent \eq{f_max}.}
\label{fig:02}
\end{figure}

In the case when monomers are separated by a significant contour distance, i.e., when $s_1 = |n-k| \gg 1$ and $s_2 =|m-n| \gg 1$, only the eigenvalues with small $p$ ($p \ll N$) are important. Moreover, it is possible to further distinguish between the cases of $s_{1,2} \sim N$ and $s_{1,2} \ll N$. In the first case, only several first eigenvalues are relevant and the behavior is not universal, i.e., for example, the behavior of linear chains and cycles is different. Meanwhile, for $1 \ll s_{1,2} \ll N$ it is not just several first eigenvalues that are important but the whole shape of the edge of the spectrum. In particular, if for $p \ll N$ the spectrum has the form $\omega_p \sim (p/N)^{2/d_s}$, where $d_s$ is called the
spectral dimension of the elastic graph \cite{dolgushev1}, then the resulting equilibrium conformation is self-similar with monomer-monomer distance distribution given by \eq{fGaussian} and fractal dimension $d_f$ satisfying
\be
2^{-1} = d_s^{-1} - d_f^{-1}
\ee
It is exactly what happens in the case of the Hamiltonian \eq{quadr} with power-law decaying coefficients \eq{decay} (see \cite{polovnikov_sm18}). This allows one to use \eq{fd} and \eq{fd2} for the mean-square distance between the monomers, and distance-distance correlations, respectively, resulting in the following expression for the correlator $r(k,n,m)$, which, as it turns out, depends in this case on a single variable $\chi = s_1/s_2= (n-k)/(m-n)$
\be
r(\chi) = \frac{1}{2}\left(\left(\chi^{1/2}+\chi^{-1/2}\right)^\alpha-\chi^{\alpha/2}-\chi^{-\alpha/2} \right); \quad \alpha = 2/d_f
\label{rt}
\ee
As one would expect, the correlation function $r(\chi)$ is symmetric $r(\chi) = r(1/\chi)$ and negative for all non-zero $\chi$ for the subdiffusive fractional Brownian motion ($d_f>2$). If $d_f=2$, $r(\chi) \equiv 0$, as it should be for normal Brownian motion. Figure \fig{fig:02} left shows the dependence $r(\chi)$. It is clearly seen that correlation is the strongest (the absolute value of coefficient is the largest) for the loops of equal size, the corresponding value being $r(1) = 2^{\alpha-1}-1$. If the loops lengths are very different $\chi \gg 1$ the correlation coefficient converges to zero as a power law
\be
r(\chi) \sim - \frac{1}{2} \chi^{-\alpha/2} ; \quad \chi \gg 1
\ee

Summing up, the total probability of a triple contact in a case of a fractional Brownian motion trajectory reads
\be
P_3(s_1,s_2) = \left(\frac{3}{2\pi b^2 }\right)^{3} (s_1 s_2)^{-3/d_f} f(s_1,s_2) = P_3^{\text{naive}} \times f(s_1,s_2),
\label{triple_2}
\ee
where
\be
f(s_1,s_2) = (1-r^2(s_1/s_2))^{-3/2};
\label{triple_f2}
\ee
$s_1, s_2$ are contour lengths of the loops, $b$ is a characteristic microscopic length, and
\be
P_3^{\text{naive}} = \left(\frac{3}{2\pi b^2 }\right)^{3} (s_1 s_2)^{-3/d_f}
\ee
is a product of single loop probabilities. Note that the naive guess always underestimates the loop probability (because it neglects the effective attraction of the chain units). The largest boost in the loop probability is achieved for the loops of equal size and equals
\be
f_{\max}(\alpha)=2^{-3\alpha/2} (1-2^{\alpha-2} )^{-3/2},
\label{f_max}
\ee
meaning, e.g., that in the case $\alpha =2/3$ ($d_f=3$), which is the most interesting case from the polymer physics point of view, the number of rosettes consisting of two loops of equal size should be some 7\% larger than one would expect from the naive uncorrelated loop assumption. Notably, this boost increases rather rapidly with the fractal dimension (\fig{fig:02} right), reaching, e.g., roughly 14\% for $d_f=4$, which is reported as a transient fractal dimension in some polymer systems, e.g., ideal randomly branched polymers \cite{stockmayer,daoud_joanny,everaers_grosberg}. In turn, for the triple rosettes formed with loops of very different size one expects the boost factor
\be
f(\chi) =
1 + \frac{3}{8}\chi^{-\alpha} + o\left(\chi^{-\alpha}\right)
\label{large_chi}
\ee
We expect that with the development of many-body Hi-C techniques it will become possible to experimentally measure this excess formation of triple contacts as compared to the product of the probabilities of two independent loops. This prediction is characteristic of the Gaussian model we use here: indeed, in the system with hard-core short-range interactions one expects the formation of triple loops to be suppressed, not boosted. Thus, if it is indeed observed experimentally, it would be a nice new argument in support of the Gaussian formalism. Also, equations \eq{f_max} and \eq{large_chi} allow new alternative ways of measuring $\alpha$ and, therefore, the effective fractal dimension $d_f$ of chromatin packing, which can be further compared to the results obtained by different methods.


In conclusion, let us briefly discuss the generalization of the formalism developed above for the calculation of the probabilities of many-body contacts. The probability $P_k$ for a polymer interacting with itself via the potential \eq{quadr} to form a $(k-1)$-rosette made of loops anchored by a set of monomers with numbers $\{n_1, n_2, ..., n_k\}$ can be written as
\be
\begin{array}{rll}
P_k(n_1, ..., n_k) &=& \dfrac{\displaystyle \int P(\ve X) \prod_{i=1}^{k-1} \delta(\ve y_i) D\ve X}{\displaystyle \int P(\ve X) D\ve X }=
\medskip \\
&=& \displaystyle
\frac{1}{(2\pi)^{3(k-1)}} \int \prod_{i=1}^{k-1} d\mathbf{q}_i
\exp\left(-\sum_{p} \frac{1}{4N\kappa_p} \left|\sum_{i=1}^{k-1}\mathbf{q}_i \left(e^{\frac{2i\pi p n_i}{N}}-e^{\frac{2i\pi p n_{i+1}}{N}}\right)\right|^2\right),
\end{array}
\label{many}
\ee
where we introduced the notation $\mathbf{y}_i = \mathbf{x}_{n_i} - \mathbf{x}_{n_{i+1}}$.
Proceeding in the same way as above one can express the integral on the right hand side of \eq{many} in the following way, which is direct generalization of \eq{sigma}:
\be
P_k(n_1, ..., n_k) = \left(\frac{1}{2\pi}\right)^{3(k-1)/2} \left(\det \Sigma_{k-1}\right)^{-3/2},
\label{many_res}
\ee
where $\Sigma_{k-1}$ ia a covariance matrix of vectors connecting successive monomers in the cluster: $\left(\Sigma_{k-1}\right)_{ij} = \langle \ve y_i \ve y_j \rangle$.
These expressions can be further simplified, for example for the probability of 4-body contacts one gets
\be
P_4(n_1,...n_4)=P_2(n_1,n_2) P_2(n_2,n_3) P_2(n_3,n_4) f_3(n_1,n_2,n_3,n_4),
\ee
where the function $f_3$ equals
\be
f_3(n_1,n_2,n_3,n_4) = \left( 1- r_{12}^2 - r_{23}^2 - r_{13}^2 + 2 r_{12} r_{23} r_{13}\right)^{-3/2};\quad r_{ij} = \frac{\langle \ve y_i \ve y_j \rangle}{\sqrt{\langle \ve y_i^2 \rangle\langle \ve y_j^2 \rangle}}.
\ee




Summing up, the main results of our work are the formulae \eq{triple_1}, \eq{triple_f} for the triple contacts, and \eq{many_res} for multiple contacts in a Gaussian polymer with Hamiltonian \eq{quadr}. It is known that for long polymer chains with interaction coefficients decaying as \eq{decay} the conformations of such polymers are statistically the same as trajectories of fractional Brownian motion, leading to the result \eq{rt} for the correlation coefficient, and formulae \eq{triple_2} and \eq{triple_f2} for three-body contacts. We believe that these results constitute a basic benchmark for interpretation of the experimental data on many-body contacts in fractal polymer conformations with $d_f>2$, in particular, the packing of chromosomes, which we expect to be available soon. On the other hand, from the random walk theory point of view, we expect our result to be essential for the understanding of the localization phenomena of random walks with long-ranged memory, e.g., adsorption of such a walker on a point-like potential well.

\section*{Acknowledgements}
The authors are grateful to L. Mirny for stimulating discussions on the topic of this work. SN is grateful to the RFBR grant 18-29-13013mk for partial support, KP and MT acknowledge the support of the Foundation for the Support of Theoretical Physics and Mathematics ``BASIS'' (grant 17-12-278).

\end{document}